\documentclass[conference]{IEEEtran}
\IEEEoverridecommandlockouts
\usepackage{cite}
\usepackage{amsmath,amssymb,amsfonts}
\usepackage{algorithmic}
\usepackage{graphicx}
\usepackage{textcomp}
\usepackage{xcolor}
\def\BibTeX{{\rm B\kern-.05em{\sc i\kern-.025em b}\kern-.08em
    T\kern-.1667em\lower.7ex\hbox{E}\kern-.125emX}}

\newtheorem{definition}{Definition}
\usepackage{subfigure}        
\begin{document}

\title{Adaptive Differential Privacy in Federated Learning: A Priority-Based Approach}

\author{\IEEEauthorblockN{1\textsuperscript{st} Mahtab Talaei}
\IEEEauthorblockA{\textit{Department of Electrical and Computer Engineering} \\
\textit{Isfahan University of Technology}\\
Isfahan, Iran \\
mtalaei@bu.edu\textsuperscript{1}}
\and
\IEEEauthorblockN{2\textsuperscript{nd} Iman Izadi}
\IEEEauthorblockA{\textit{Department of Electrical and Computer Engineering} \\
\textit{Isfahan University of Technology}\\
Isfahan, Iran \\
iman.izadi@iut.ac.ir}
}

\maketitle

\begin{abstract}
Federated learning (FL) as one of the novel branches of distributed machine learning (ML), develops global models through a private procedure without direct access to local datasets. However, access to model updates (e.g. gradient updates in deep neural networks) transferred between clients and servers can reveal sensitive information to adversaries. Differential privacy (DP) offers a framework that gives a privacy guarantee by adding certain amounts of noise to parameters. This approach, although being effective in terms of privacy, adversely affects model performance due to noise involvement. Hence, it is always needed to find a balance between noise injection and the sacrificed accuracy. To address this challenge, we propose adaptive noise addition in FL which decides the value of injected noise based on features' relative importance. Here, we first propose two effective methods for prioritizing features in deep neural network models and then perturb models' weights based on this information. Specifically, we try to figure out whether the idea of adding more noise to less important parameters and less noise to more important parameters can effectively save the model accuracy while preserving privacy. Our experiments confirm this statement under some conditions. The amount of noise injected, the proportion of parameters involved, and the number of global iterations can significantly change the output. While a careful choice of parameters by considering the properties of datasets can improve privacy without intense loss of accuracy, a bad choice can make the model performance worse.
\end{abstract}


\footnote[1]{Mahtab Talaei was affiliated with the Department of Electrical and Computer Engineering of Isfahan University of Technology, during the research for this paper. Her current affiliation is with the Division of Systems Engineering at Boston University, Boston, US.}

\section{Introduction}
With the development of computational and communicational capabilities of distributed systems, including smart phones, sensor networks, Internet-of-Things (IoT), and the rapid growth of the applications of these systems in our daily lives, an unprecedented amount of data is being produced everyday~\cite{bib1}. Therefore, utilizing these sources of rich information to upgrade features and services offered to people and organizations owning this data, without violating their privacy, is of great significance. Distributed machine learning (ML) is a promising solution in settings dealing with large volumes of data as well as privacy concerns about users' sensitive information leakage. 

Due to an increasing emphasis on users' privacy, federated machine learning techniques are widely exploited, and global models are developed by the use of local datasets available only on each client~\cite{bib2, bib3, bib13}. While offering many advantages over conventional machine learning methods, federated learning (FL) has its own challenges, including expensive communication costs, systems and statistical heterogeneities, and privacy concerns~\cite{bib5, bib2, bib14, bib16, bib19}.

Deep learning (DL) is one of the most popular algorithms used in the context of ML, especially when we are willing to extract features from large image, voice, or text datasets~\cite{bib6}. Therefore, FL can benefit from these algorithms while developing artificial intelligence (AI) models. In order to optimize FL local models, the stochastic gradient descent (SGD) method is generally adopted~\cite{bib12}.

Preserving users' privacy and data security is perhaps one of the most debatable topics in all ML algorithms. Even though the idea of federated learning was first proposed for its better privacy guarantees, several experiments have shown that detecting users' private data is still possible from the gradient updates sent from clients to the server~\cite{bib20}. Especially, when DL models are designed for local models, the risk of revealing training data by the access to developed models increases~\cite{bib26}. Besides using cryptographic methods, such as secure multi-party computations (MPC) and homomorphic encryption (HE) schemes~\cite{bib21}, differential privacy (DP) is widely used in FL for data protection.

DP tries to reduce the risk of information leakage by adding deliberate noise to data. Laplace, exponential, and Gaussian mechanisms are three fundamental noise injection mechanisms for implementing DP~\cite{bib23}.~\cite{bib12} presented a framework for global differential privacy, which accurately calculates the Gaussian noise required for DP. It also gives a theoretical explanation for the convergence behavior of the suggested scheme.~\cite{bib22} proved that high privacy DL models can be developed in distributed systems using selective sharing of small subsets of local key parameters. They showed that sharing only $10$ percent of local parameters results in better models than non-collaborative models. They also applied differential privacy by perturbing the gradients of the proportion of local parameters selected for sharing. However, perturbing model parameters and injecting noise have a definite consequence: accuracy loss!

Most of the existing works on DP of federated systems inject a constant amount of noise into all parameters. This approach, although protects data, impacts the model performance. Since each parameter doesn't have an identical effect on the model's output, it seems that adding noise based on this difference can enhance the developed model performance. This idea may be able to address the challenge of losing accuracy in exchange for a better privacy in DP algorithms, to some extent. To the authors' knowledge, the only work on adaptive DP is~\cite{bib29}, which injects Laplace noise based on the contributions of neurons to the output. The contributions of input features are calculated from the contributions of the next layer neurons.This backward scenario, which is repeated until finding the input features' contributions, can be both time-consuming and computationally expensive for clients, considering a large number of iterations and high-dimensional weight matrices. Consequently, the adaptive noise is injected to the contributions themselves. 

On  the other extreme, there are numerous works with the terms ``adaptive" and ``personalized" DP~\cite{bib30, bib31}, which choose fixed noise parameters for each client based on the whole local database it has. These parameters may differ from one device to the other, but remains the same for all the parameters of each local model. The main concern of adaptive clipping or personalized DP methods is statistical heterogeneities between the clients.

In this paper, however, we propose an adaptive differential privacy framework with Gaussian noise for deep learning models developed in federated learning. This adaptive approach is based on features' and parameters' importance. In fact, adding noise to less important features does not affect accuracy as adding noise to important features does, and at the same time, it can preserve privacy. So, we first present a solution for prioritizing neurons, which can depend on the inputs and outputs only, and then, try to distinguish the relations between perturbing important and unimportant features and the accuracy by simulations. Our main concern is to attain a balance between accuracy and privacy.

Although for an adaptive DP, it is often assumed that more noise should always be added to irrelevant features, our studies show a contrary result. The amount of noise, the proportion of important and unimportant features involved, and the number of global iterations can have significant effects on the outcome. When we add noise to $20$ percent of parameters with high importance, we observe a noticeable reduction in accuracy compared with when the same amount of noise is added to $20$ percent of low importance parameters. However, when a larger number of parameters are involved in the adaptive approach, the outcome is not that straightforward. Generally, there is a trade-off between noise and accuracy. With a reasonable compromise, we can preserve privacy without a significant accuracy loss, but otherwise, it can ruin model performance.

The remainder of this paper is organized as follows. In Section \ref{two}, we review some preliminaries on DL, FL, and DP. In Section \ref{three}, we introduce our approach for making DP adaptive in FL. The experimental evaluations and results are presented in Section \ref{four}, and the summary and conclusion are given in Section \ref{five}.

\section{Preliminaries} \label{two}
In this section, we review some key materials for DL, FL, and DP.

\subsection{Deep Learning}
\begin{figure}[b]
\begin{center}
\includegraphics[width=0.8\linewidth]{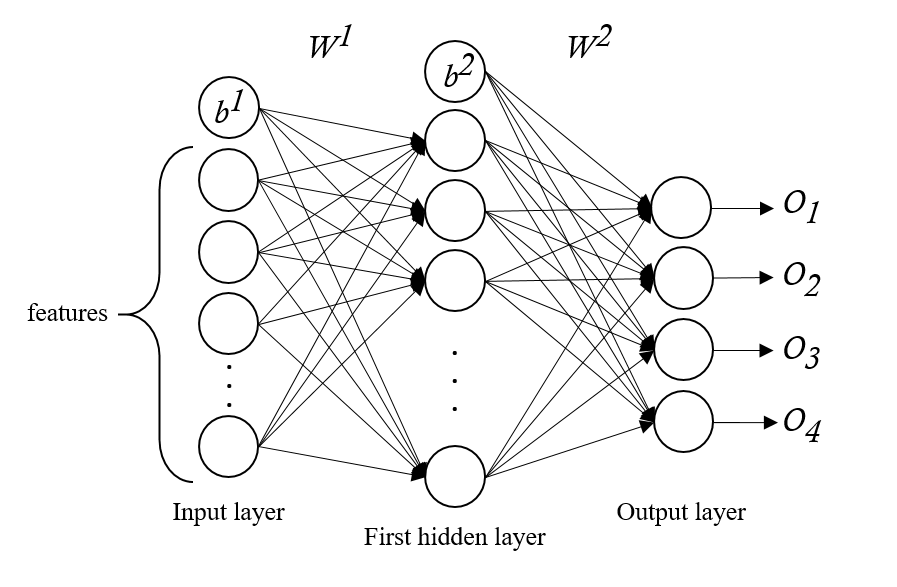}
\end{center}
\caption{A simple neural network with one hidden layer. $b^k$ vectors contain the bias values for each neuron. 
$W^k$ is the weight matrix of each layer.}\label{fig:NN}
\end{figure}

Deep learning is an AI function that tries to extract important features of big data by imitating the work of the human brain. Neural networks as a subset of deep learning models have a web-like multi-layer structure through which inputs are connected to outputs~\cite{bib28}. Fig.~\ref{fig:NN} depicts a simple neural network with one hidden layer. Nodes in each layer, known as neurons, are connected to the next layer neurons via weight vectors, which are updated through the learning process to perfectly extract the relations between inputs and outputs. Each neuron output is calculated by a non-linear activation function $f( W^k, a^{k-1}, b^k)$, where $W^k$ is the weight matrix of the $k$-th layer, $a^{k-1}$ is the previous layer neuron outputs and $b^k$ is the bias vector.

Similar to all the other ML algorithms, the goal in neural networks is to find the weights matrix minimizing the defined loss function. Generally, in supervised learning, random weights will be first considered and the output will be found by these parameters through the feed-forward procedure. After calculating the error between the computed and expected outputs, gradient descent in the back-propagation mode will be used in order to update the weights of the network. Since using gradient descent for all data samples in large datasets seems irrational, SGD, which operates over randomly chosen smaller subsets of datasets, is an alternative. If we transform the weights matrix to a vector and let $E_i$ be the loss function over the mini-batch $i$, the updated weights are calculated by
\begin{equation}
w_j = w_{j} - \alpha \frac{\partial E_i}{\partial w_j},
\end{equation}
where $\alpha$ is the learning rate controlling the step size. 

Generally, neurons and weights in the lower layers have stronger effects on learning~\cite{bib22}. For instance, in image processing, raw pixels of a picture are the first layer neurons, and hence, the parameters directly in touch with them are the most effective parameters for the value of the network output. We further use this observation for feature importance rating.

\subsection{Federated Learning}
\begin{figure}[b]
\begin{center}
\includegraphics[width=0.8\linewidth]{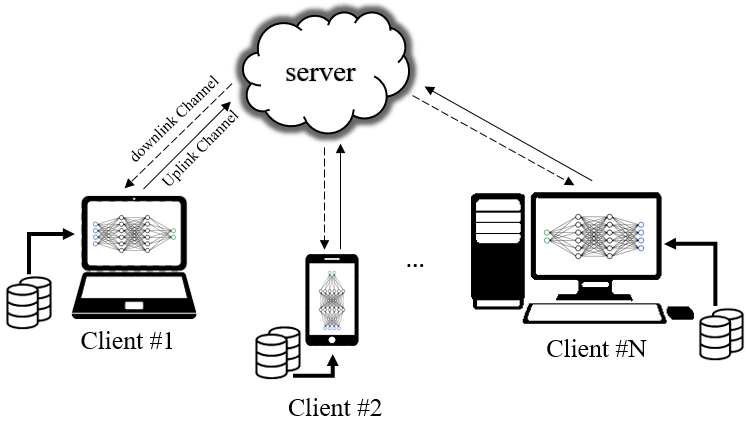}
\end{center}
\caption{FL training model. }\label{fig:FL}
\end{figure}

The goal in a standard federated learning problem is to develop a global ML model for tens to millions of clients without direct access to their local data~\cite{bib5}. The only messages transmitted from the clients to the server, in this framework, are the training parameter updates of the local objective functions. To formalize this goal, consider $N$ clients as depicted in Fig.~\ref{fig:FL}. We wish to minimize the following loss function:  
\begin{equation}
\min_w F(w),\, \textnormal{where} \, F(w):= \sum _{i=1} ^N p_i F_i(w)
\end{equation}

Here, $F_i$ is the local loss function for the $i$-th client and the parameter $p_i$ is defined based on the relative impact of each client. Let $p_i > 0$ and $\Sigma _ i p_i=1$. The impact factor $p_i$ can be defined by $\frac{n_i}{n}$, where $n_i$ is the number of data samples of client $i$ and $n=\Sigma_i n_i$ is the total number of data samples available. The outline of the training process of FL is as the following steps~\cite{bib13, bib4}:
\begin{enumerate}
\item A central server sends a primary model to all or a subset of clients selected for the training.
\item The clients update model parameters using their local data and send the ML parameters to the server.
\item The server aggregates the received parameters using a defined algorithm, such as weighted averaging. 
\item The server returns the updated global parameters to the selected clients for another iteration until the acceptable accuracy is acquired or a sufficient number of iterations is completed.
\end{enumerate}

We assume a trustworthy server and hence, based on the aforementioned FL procedure, information leakage can only happen during transferring model parameters from the clients to the server and vice versa. Privacy protection in this framework is defined as global privacy~\cite{bib3}. 

\subsection{Differential Privacy}
DP gives strong guarantees to preserve data in ML algorithms. A randomized mechanism $\mathcal{M}$ is supposed to be differentially private if its output is robust to any change of one sample in input data. The following definition formally clarifies this statement for $(\epsilon, \delta)$-DP~\cite{bib23}:

\begin{definition}[$(\epsilon, \delta)$-DP]
A mechanism $\mathcal{M}$ satisfies $(\epsilon, \delta)$-differential privacy for two non-negative numbers 
$\epsilon$ and $\delta$ if for all adjacent datasets $\mathcal{D}$ and $\mathcal{D}^{\prime}$ such that $d(\mathcal{D},\mathcal{D}^{\prime}) = 1$, and all subsets $S$ of
$\mathcal{M}$'s range, there holds
\begin{equation}
\textnormal{Pr}[\mathcal{M}(\mathcal{D}) \in S] \leqslant e ^ \epsilon \textnormal{Pr}[\mathcal{M}(\mathcal{D}^{\prime}) \in S] + \delta
\end{equation}
\end{definition}
Here, the difference between two datasets $\mathcal{D}$ and $\mathcal{D}^{\prime}$, $d(\mathcal{D},\mathcal{D}^{\prime})$, is typically defined as the number of records on which they differ.

It is concluded from this definition that, with a probability of $\delta$, the output of a differentially private mechanism on two adjacent datasets varies more than a factor of $e ^\epsilon$. Thus, smaller values of $\delta$ enhance the probability of having the same outputs. Smaller values of $\epsilon$ narrow the bound for privacy protection. The smaller $\epsilon$ and $\delta$ are, the lower the risk of privacy violation. 

Based on ~\cite{bib23}, considering $s$ as an arbitrary $d$-dimensional function applied on a dataset, for $\epsilon \in (0,1)$ and $c \geqslant \sqrt{2 \ln{(1.25/ \delta)}}$, a Gaussian mechanism with parameter $\sigma \geqslant c \Delta s / \epsilon $ that adds noise scaled to $\mathcal{N}(0,\sigma ^ 2)$ to each component of the output of $s$ is $(\epsilon, \delta)$-differentially private. Here, $\Delta s$ is the sensitivity of the function $s$ defined by $\Delta s = \max _{\mathcal{D}, \mathcal{D}^{\prime}} \lVert s(\mathcal{D}) - s({\mathcal{D}^{\prime}}) \rVert $.

\section{Federated learning with adaptive differential privacy } \label{three}
In this section, we first briefly discuss the DP algorithm presented in ~\cite{bib12} and then try to make it adaptive based on feature and parameter importance in each client.
Here, As we cannot use those straightforward feature selection techniques offered for other ML methods in DL algorithms due to its black box structure, we tried our best
to propose two practical ways for computing features' relative importance for FL.

\subsection{Federated Learning with Differential Privacy}
We first consider all clients to participate in learning. For each global iteration, neural network weights developed locally are sent to the server for aggregation. Based on the threat model and the analysis in ~\cite{bib12}, in order to ensure $(\epsilon, \delta)$-DP for uplink parameter transmission, the Gaussian noise should be added to every single neuron-to-neuron weight with the noise distribution $\mathcal{N}(0,\sigma_U ^ 2)$, where
\begin{equation}
\sigma_U= c L \Delta s_U/ \epsilon \label{eq:4}
\end{equation}
Here, $L$ is the maximum number of exposures of the local parameters during uplink transmission. $s_U$ for each client is the output of the local training process of that client, $w_i$. So, the maximum possible value of the sensitivity of $s_U$, defined as the maximum change of $s_U$ seen by a change in one sample of the local datasets, is $\frac{2C}{m}$, with $C$ defined as the clipping parameter, $\lVert w_i \rVert \leqslant C$, and m the minimum size of local datasets.  

After perturbing local weights in the client side, the server receiving the parameters decides whether more noise should be added to the aggregated parameters for downlink transmission or not. Considering $T$ global iterations and $N$ clients, the standard deviation of the Gaussian noise added to aggregated parameters in order to ensure $(\epsilon, \delta)$-DP for downlink channel is  
\begin{equation}
\sigma_D = 
	\begin{cases}
      \frac{2cC \sqrt{T ^2 - L^2 N}}{m N \epsilon} & T > L \sqrt{N}\\
      0 & T \leqslant L \sqrt{N}
    \end{cases}
\end{equation}

\subsection{Adaptive Differential Privacy in Federated Learning}
We are interested in developing an adaptive DP framework to lower the risk of accuracy loss. Previous DP algorithms in FL, including NbAFL, proposed in~\cite{bib12}, tried to add noise with a fixed distribution to parameters to ensure global privacy. However, it seems selecting noise distribution based on the relevance of each feature to the model efficiency can satisfactorily enhance accuracy while preserving privacy. In order to evaluate this hypothesis, it is required to first prioritize parameters' importance based on their effect on the model output. 

Assume deep neural networks are used for training local models in FL. As noise should be added to the clients' parameters before uploading for the server in each iteration, we cannot use feature selection methods~\cite{bib24} that rely on the final developed models. In other words, prioritizing parameters should be done in each iteration for every client, so that the result could be used for the selection of the noise value. FL with non-IID data will also benefit from a personalized adaptive perturbing mechanism, which works in parallel for all devices. Moreover, this synchrony of local computations on the new improved and powerful distributed GPUs reduces delays in the overall training process. 

\subsection{Sensitivity-Based Method}
The first method we propose for the aforementioned parameter ranking in neural networks is based on output sensitivity. Here, after the local parameters are updated in the clients, the accuracy of the local model is calculated as a reference and saved in $acc_{ref}$. Then, an adequate value proportionate to the standard deviation of the Gaussian noise proposed in \eqref{eq:4} is chosen. Since a random variable with a Gaussian distribution with mean $\mu = 0$ and standard deviation $\sigma$ has an expected absolute value equal to $\sigma \sqrt{\frac{2}{\pi}}$, this value can be a good choice. Based on the importance of the first layer weights, we use them as our leader through this task. Hence, considering $m$ features, the selected noise value is added to all the weights connected to each input feature $W^1_p (p=1, 2, \dots,m)$ at a time. In other words, if $(W^1)^+$ denotes the new model's first layer weights matrix, $({W^1_i})^+ = W^1_j \; \,  \forall i \neq j$ and $({W^1_i})^+ = W^1_i + r$, where r is the vector with the appropriate size containing values equal to the selected noise. At the same time, the accuracy of each new $m$ model is calculated and saved in an $m$ element vector $acc^+$. The absolute difference between each of these elements and $acc_{ref}$ is defined as the relevance factor of that feature and its related weights to output. Larger output differences specify parameters more sensitive to the change. As a result, the accuracy loss employing DP on these parameters are more significant.

\subsection{Variance-Based Method}
Another approach for discovering feature importance is the variance-based method. The idea that weights connected to more important features undergo more change during the training process of neural network~\cite{bib25}, is the underlying principle for this approach. This probably is due to the fact that utilizing gradient descent for updating parameters, those parameters, having more effect on minimizing the loss function, have larger gradients. So, combing the variances of all weights connected to each input neuron through the learning process leads to a value roughly proportional to its impact on the output. Here, in the context of FL, we have several limitations. First, storing all the weights during the learning process to compute variances is computationally expensive. Moreover, clients participating in learning are not fixed and each of them can update parameters only once. Finally, we want to use the feature importance data during training for choosing appropriate DP parameters. Hence, our proposed method should work for each client in each iteration and before uplink transmission. For this sake, we assume the updated local weights are the final weights for that client and compute the variance between the new local and last global weights sent to the clients. Considering $w_{ab}$ as the weight connecting input feature $a$ to neuron $b$ of the first hidden layer, the relative feature importance $FI$ is defined as

\begin{equation}
FI_i = \sum _ {j=1} ^{q} Var(w_{ij}) \times |w_{i,j}|  \quad i=1,2, \dots ,m
\end{equation}
where $q$ is the number of neurons in the first hidden layer.

Now the question is how to use this information for adaptive $(\epsilon, \delta)$-DP. One possible method is to first choose the appropriate $\epsilon$ and $\delta$ that satisfy the least level of privacy guarantee we desire. Then, we can lower the $\epsilon$ value for the first layer parameters based on parameters importance for a stronger guarantee. A lower $\epsilon$  leads to a higher noise level. Thus, it seems the addition of more (higher values of) noise to irrelevant parameters, in comparison to more important ones, has a less adverse effect on model accuracy.

\section{Simulation Results } \label{four}
In this section, we evaluate the proposed adaptive noise addition for FL on the real-world Modified National Institute of Standards and Technology (MNIST) dataset~\cite{bib27}. The multi-layer perceptron (MLP) is employed for developing a global model and global differential privacy parameters are chosen based on the principles proposed in NbAFL~\cite{bib12}. Then, adaptive DP is applied by the use of both sensitivity and variance-based feature importance methods. These simulations are performed under different protection levels ($\epsilon$), proportions of weights chosen for more noise addition, and values of the added adaptive noise.

MNIST is a dataset for handwritten digit identification, divided into training and testing examples, which is widely used for ML experiments. Each example is a $28 \times 28$ pixel image.  A subset of 42000 images from its training set is used for this simulation with a $20$ percent split for testing. The MLP model designed for this FL task has 256 neurons in its single hidden layer with a ReLU activation function. For the output layer, a softmax of 10 classes is used. The learning rate and the clipping value for the SGD optimizer are set to $0.02$ and $5$, respectively.

\begin{figure}[h]
\begin{center}
\includegraphics[width=0.8\linewidth]{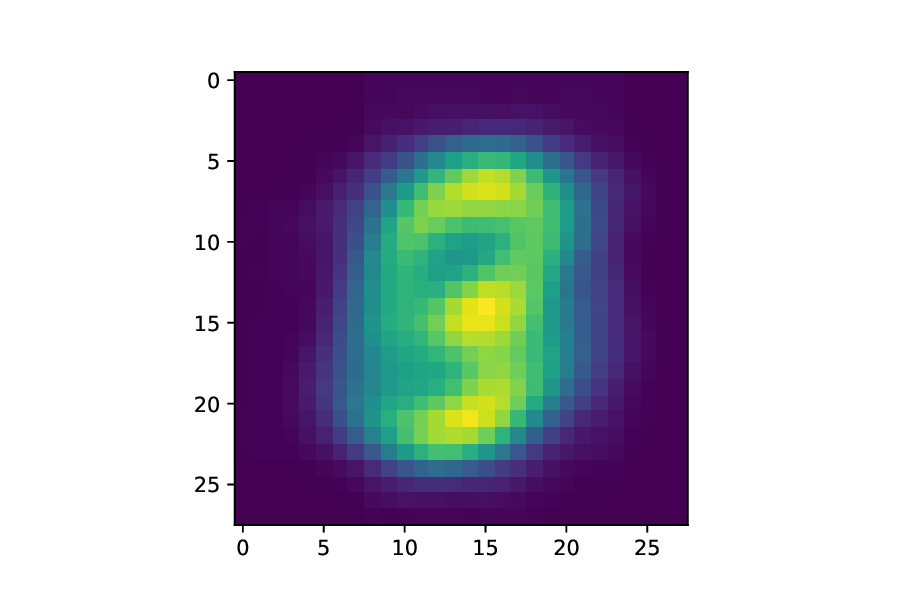}
\caption{Average of each $784$ pixels of all local samples in the first client. }\label{fig:C}
\end{center}
\end{figure}

\begin{figure}[h]
\centering     
\subfigure[Sensitivity-Based Method]{\label{fig:a1}\includegraphics[width=0.49\linewidth]{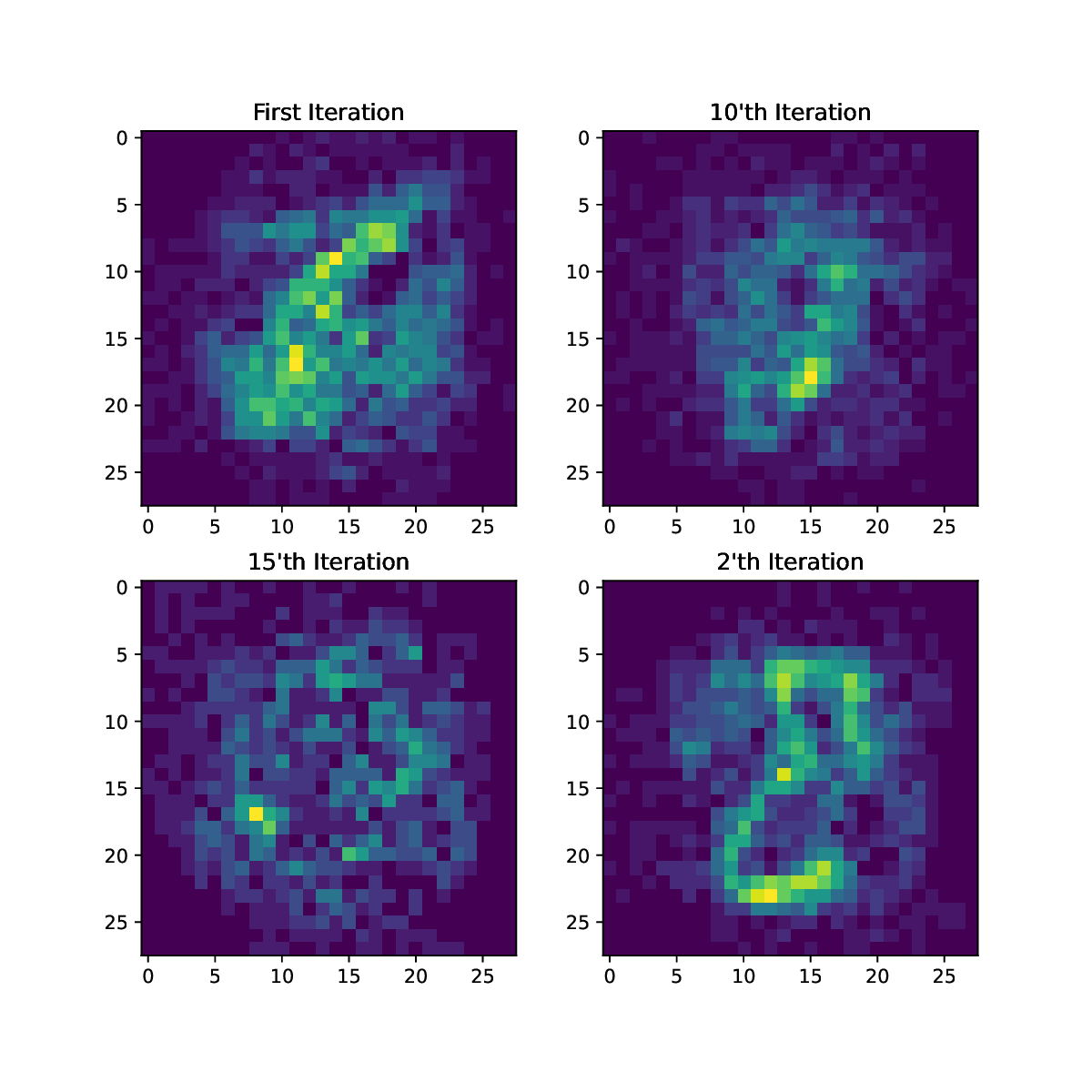}}
\subfigure[Variance-Based Method]{\label{fig:b1}\includegraphics[width=0.49\linewidth]{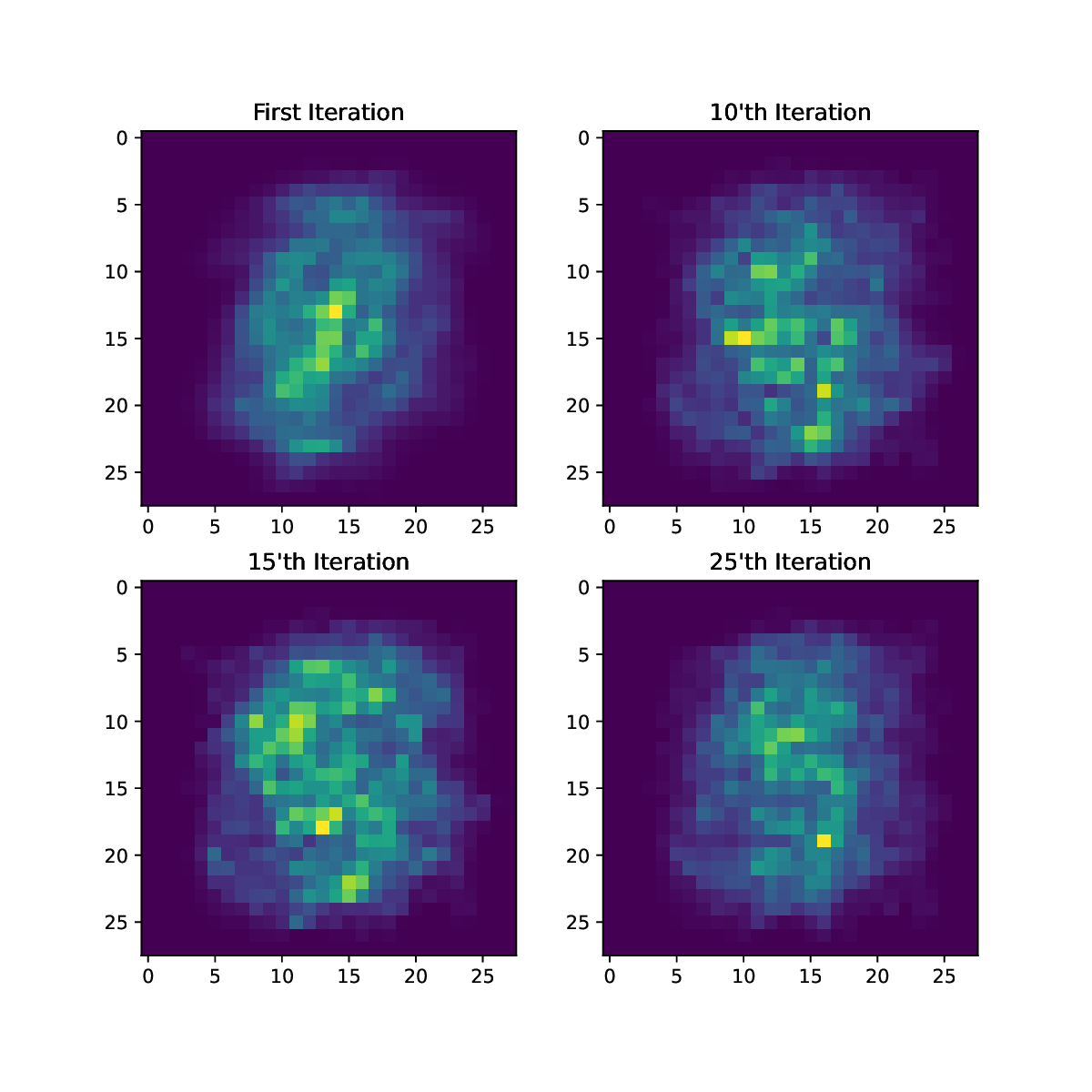}}
\caption{The relative feature importance assigned to the first client samples for four different global iterations. }\label{fig:FI}
\end{figure}

In this simulation, we set the number of clients and the number of global iterations equal to $30$ and $25$, respectively. To show the results of the proposed feature importance methods, we applied them in a non-private FL setting. For a client with an average of local samples depicted in Fig.~\ref{fig:C}, the output of four iterations of sensitivity and variance-based feature importance algorithms is shown in Fig.~\ref{fig:a1} and~\ref{fig:b1}, respectively. The darker spots identify the most irrelevant features, and as expected, unlike central points, the outer region of the handwritten digit images do not have a significant effect on predicting their class. 

\begin{figure}[t]
\begin{center}
\includegraphics[width=1\linewidth]{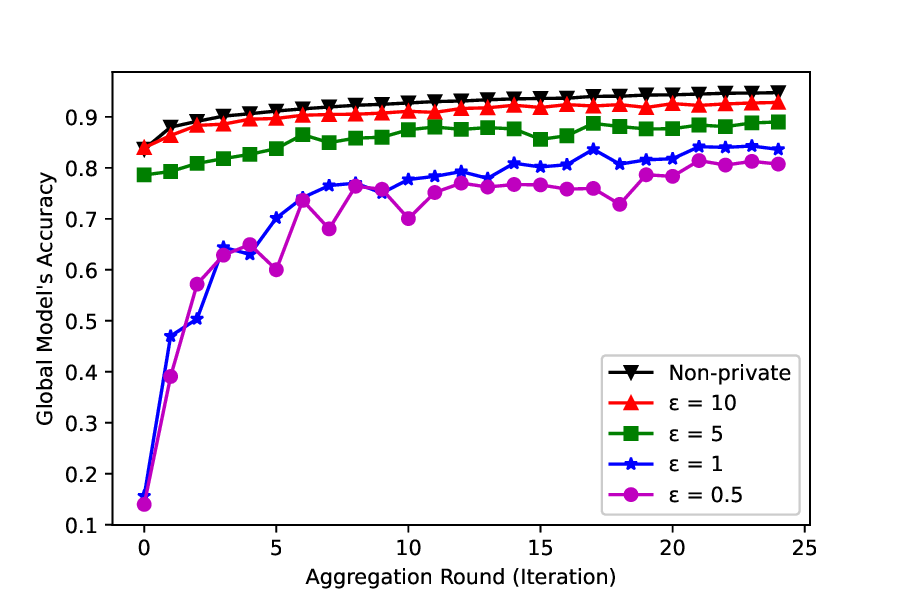}
\end{center}
\caption{The comparison of model accuracies with different protection levels $\epsilon = 0.5$, $\epsilon = 1$,
$\epsilon = 5$ and $\epsilon = 10$, for 30 clients. The additive Gaussian noise distribution is considered the same for all of the parameters.}\label{fig:acc1}
\end{figure}

Here, we first present the results of applying normal DP for various noise parameter $\epsilon$ in Fig.~\ref{fig:acc1}. We also include a non-private result for comparison. With a fixed $\sigma = 0.01$, lower $\epsilon$ values bring a better protection level in exchange for diminishing accuracy. The noise here is added to both layers' parameters with the same distribution. It is observed from this figure that $\epsilon > 1$ does not affect the quality of the model in comparison with $\epsilon$ values equal or lower than $1$.

\begin{figure}[b]
\begin{center}
\includegraphics[width=1\linewidth]{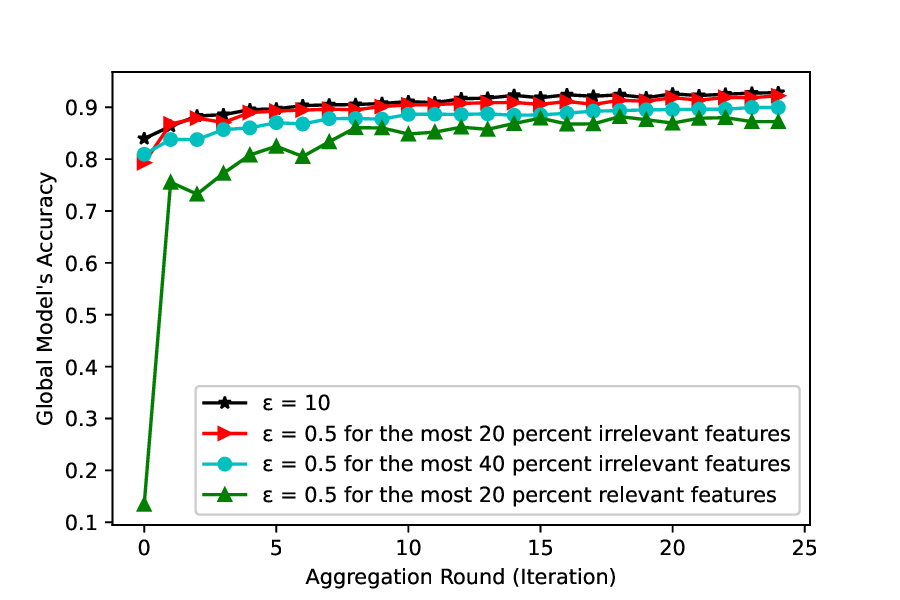}
\end{center}
\caption{The comparison of model accuracies after applying adaptive noise for the least and the most important features.}\label{fig:acc2}
\end{figure}

Fig.~\ref{fig:acc2} compares the model performance after adding adaptive noise for the most and the least important features. Using the variance-based method, a relative feature importance value is assigned to each input. $20$ and $40$ percent of the whole features with the lowest importance values and $20$ percent of them with the highest values are chosen for a protection level of $\epsilon=0.5$ while setting $\epsilon = 10$ for all of the other parameters (including the second layer parameters). The proportion of the weights of the network directly connected to $20$ percent 
of features to the whole network weights (including biases) is actually $(\frac{20}{100} \times 784 \times 256) / (785 \times 256 + 257 *10) \thickapprox 0.19$. 

As Fig.~\ref{fig:acc2} shows, although the number of parameters receiving more noise is equal for the green and red plots, there is a profound difference between the accuracy achieved. In fact, even having additional noise for twice more parameters of the least important features (the blue plot) doesn't have an impact as adverse as adding more noise on the top $20$ percent. This property can be used to achieve a better privacy level without loss of accuracy. In fact, we can find and preserve the most important parameters and perturb others. Fig.~\ref{fig:acc2} also proves the validity of our feature prioritization methods. 

\begin{figure}[t]
\centering     
\subfigure[]{\label{fig:a2}\includegraphics[width=0.49\linewidth]{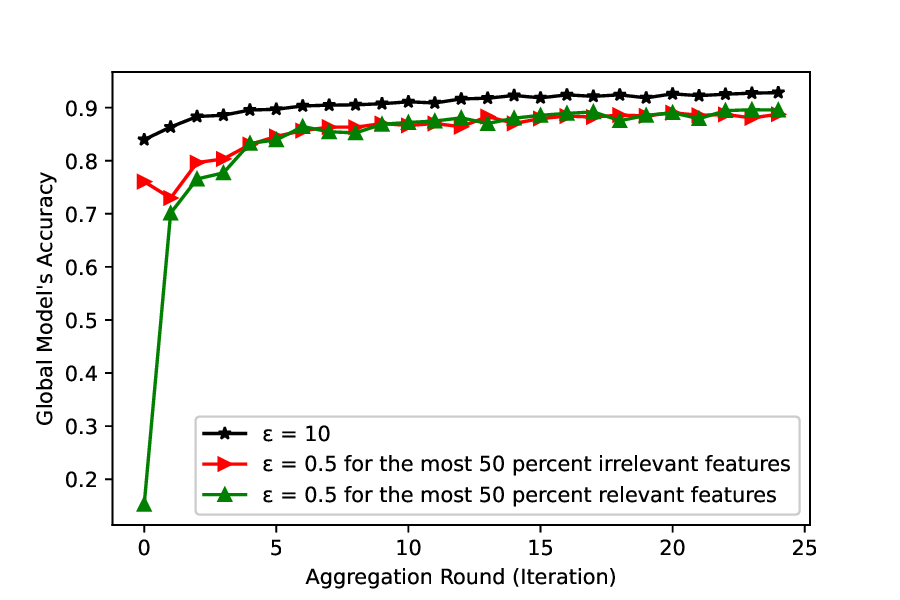}}
\subfigure[]{\label{fig:b2}\includegraphics[width=0.49\linewidth]{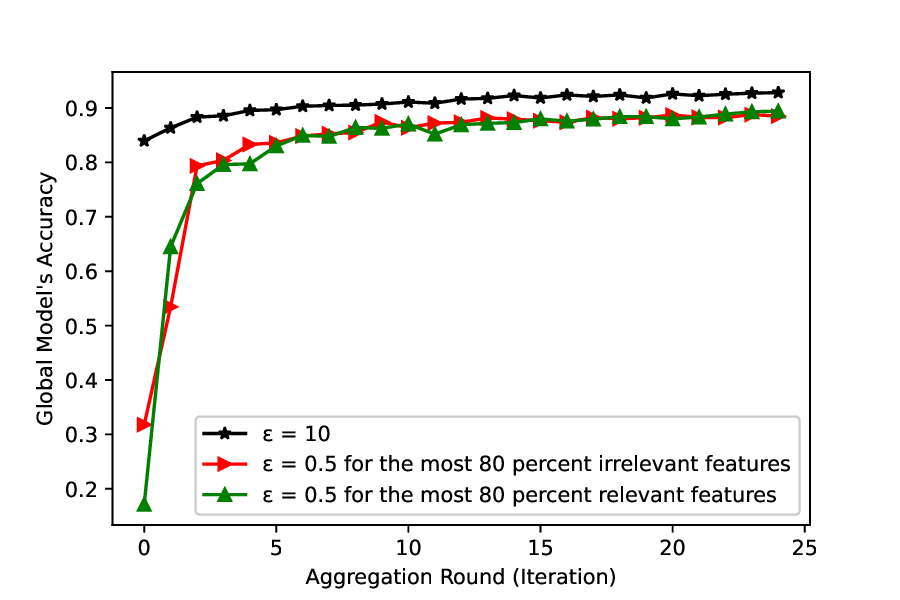}}

\caption{Comparison of model accuracies after applying adaptive noise for the least and the most important features. }\label{fig:acc3}
\end{figure}

We now increase the number of parameters involved in the adaptive approach to observe their effect on the results. Fig.~\ref{fig:acc3} represents the impact of additional noise on $50$ and $80$ percent of the first layer parameters, respectively. As the number of selected parameters increases, the importance of selective DP and prioritizing parameters diminishes. This can be due to the fact that the MNIST dataset has few important features. In other words, a small number of image pixels are informative and by increasing the features involved, noise shifts to be the decisive factor and deteriorates the model performance. 

Moreover, the difference between accuracies is more obvious at several first iterations and as time passes, the worse model improves itself and it is even possible that this model's accuracy overtakes the other one as in Fig.~\ref{fig:a2}. Neural network shows outstanding ability in enhancing its performance even in very noisy environments. Once it is properly designed, it is fairly robust. So, it can be a good choice for DP approaches. Comparing Fig.~\ref{fig:acc2} and Fig.~\ref{fig:acc3} emphasizes the necessity of choosing appropriate proportions and noise distributions for adaptive DP.

\begin{figure}[t]
\centering     
\subfigure[]{\label{fig:a3}\includegraphics[width=0.49\linewidth]{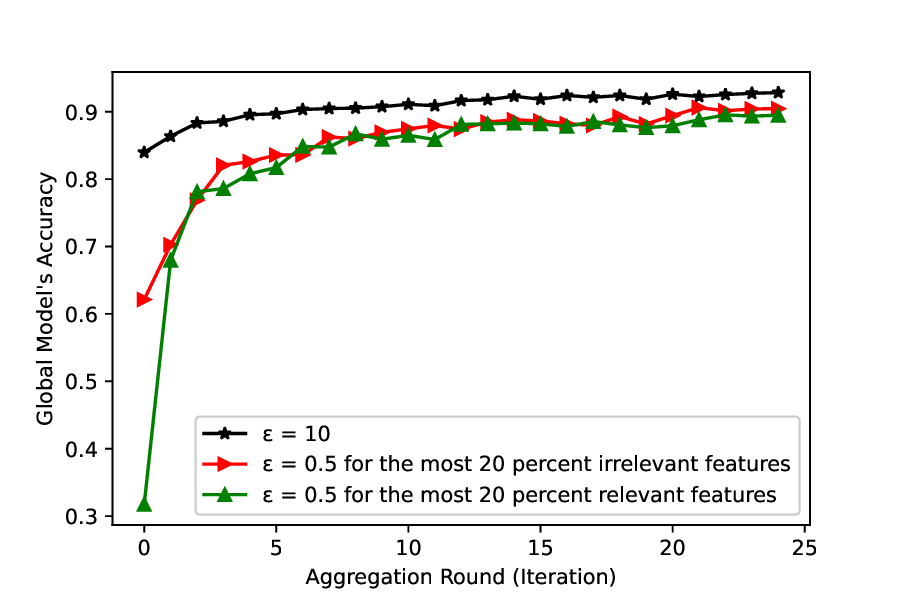}}
\subfigure[]{\label{fig:b3}\includegraphics[width=0.49\linewidth]{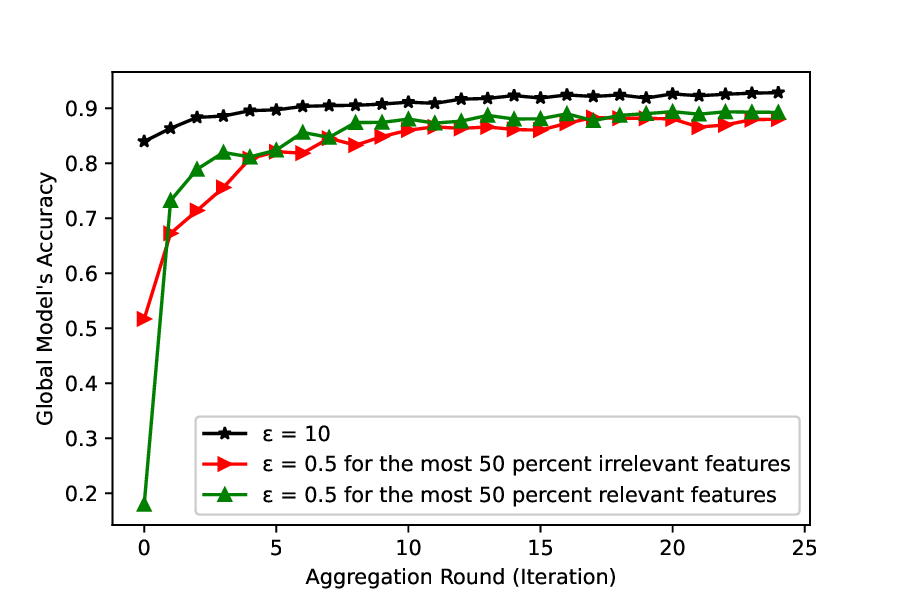}}
\subfigure[]{\label{fig:c3}\includegraphics[width=0.49\linewidth]{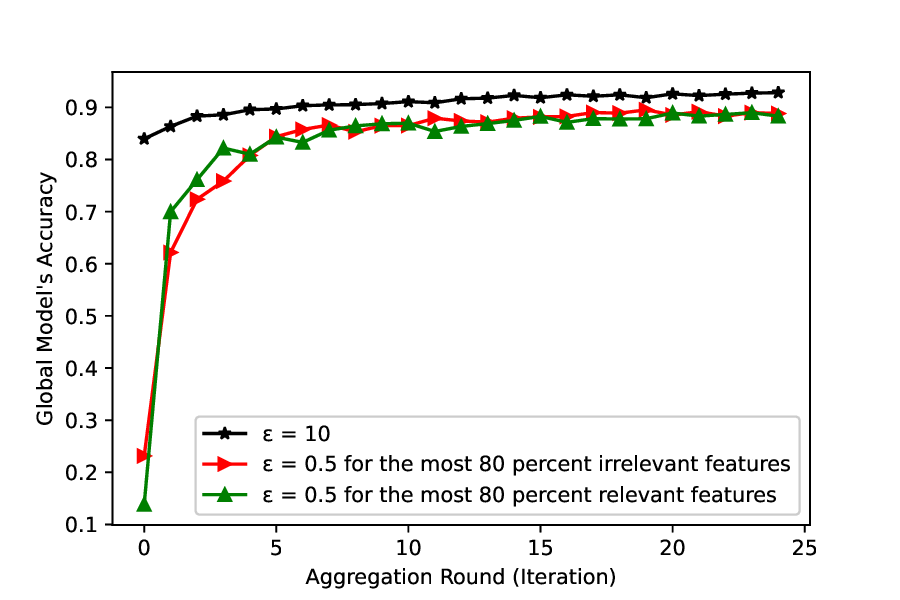}}

\caption{Comparison of model accuracies after applying adaptive noise for the least and the most important features using sensitivity-based method. }\label{fig:acc4}
\end{figure}

As Fig.~\ref{fig:acc4} depicts, repeating the same procedure with the sensitivity-based approach yields to similar overall results. The slight differences seen between the results of the sensitivity and the variance methods, especially between Fig.~\ref{fig:a3} and Fig.~\ref{fig:acc2}, states that the variance-based feature importance method prioritizes features more precisely. A more accurate feature importance method leads to more accurate results while applying adaptive DP, as variance-based method did.

\section{Conclusions} \label{five}
In this paper, we presented an adaptive differential privacy framework in federated learning based on prioritizing parameters in deep neural networks. It's proved that selective noise injection can make a profound difference in the accuracy of the global model. As it is shown, adding the same amount of noise to different parts and proportions of data, although brings the same overall privacy level, affects the model performance differently. In each simulation result, we compared various aspects of the curves from the best to the worst accuracy. These results not only confirmed the validity of the two proposed prioritizing techniques,The sensitivity-based and variance-based approaches, but also can open up a new area of research for optimizing DP in FL.  

For future possible research directions, the effect of intentionally adding irrelevant features between other features on protecting privacy should be checked. Perturbing these irrelevant features is also an alternative. Another approach is to choose noise distribution based on each parameter importance, rather than applying a fixed distribution for a proportion of the features. 
 
\bibliographystyle{IEEEtran}
\bibliography{IEEEabrv,mybibfile}

\end{document}